\documentclass{iaus}
\usepackage{graphicx}

\title[Strong lensing analysis of A1689 from deep ACS images]{Strong lensing analysis of A1689 from deep ACS images}

\author[K. Sharon]%
{Keren Sharon$^1$, Tom J. Broadhurst$^1$, Narciso Benitez$^{2,3}$, Dan Coe$^{2}$, Holland Ford$^{2}$\\and ACS Science Team } 
\affiliation{$^1$School of Physics \& Astronomy, Tel-Aviv University, Israel
email: kerens@wise.tau.ac.il\\[\affilskip]
$^2$Physics and Astronomy Dept, Johns Hopkins University, Baltimore Maryland, USA\\
$^3$Instituto de Astrof\'\i sica de Andaluc\'\i a (CSIC), C/Camino Bajo de Hu\'etor, 24, Granada, 18008, Spain}

\pubyear{2004}
\volume{225}
\pagerange{1--8}
\date{?? and in revised form ??}
\jname{Impact of Gravitational Lensing on Cosmology}
\editors{Mellier, Y. \& Meylan,G. eds.}
\begin{document}

\maketitle

\begin{abstract}

 ACS observations of massive lensing clusters permit an order of
magnitude increase in the numbers of multiply-lensed background
galaxies identified behind a given cluster. We have developed a code to
take the pixels belonging to any given image and generate
counter-images with full resolution, so that multiple systems are
convincingly and exhaustively identified.  Over 130 images of 35
multiply lensed galaxies are found behind A1689, including many radial
arcs and also tiny counter-images projected on the center of mass. The
derived mass profile is found to flatten steadily towards the center,
like an NFW profile, with a mean slope
$d{\log{\Sigma}}/d{\log{r}}\approx-0.55\pm0.1$, over the range
$r<250$~kpc/h, which is somewhat steeper than predicted for such a
massive halo.  We also clearly see the expected geometric increase of
bend angles with redshift, however, given the low redshift of A1689,
$z=0.18$, the dependence on cosmological parameters is weak, but using
higher redshift clusters from our GTO program we may derive a more
competitive constraint.

\end{abstract}

\section{Introduction}

The ACS Guaranteed Time Observations (GTO) program includes deep
observations of several massive, intermediate redshift galaxy
clusters. Our goals are to determine the distribution of matter in
clusters, to place new constrains to the cosmological parameters and
also a study of the distant lensed galaxies themselves, taking
advantage of their generally large magnifications.

This work concentrates on A1689, which has the largest known Einstein
radius of all the massive lensing clusters, of approximately 50''
in radius, based on the radius of curvature of a giant low
surface-brightness arc. Relatively little work on this cluster has
been carried out with HST, and the field of WFPC2 is too small to
cover the full area interior to the Einstein radius. Although no
actual multiple images had been identified prior to this
investigation, we were confident that the exceptional depth and high
resolution of ACS would lead to the detection of many
sets of multiple images, more than possible around other clusters, by
virtue of the large Einstein radius.

See Broadhurst et al. (2004) for a comprehensive description of this work.

\section{Observations}

The deep, ACS observations of A1689 are split into 4 filters, allowing reliable photometric redshift
information. 
 In total we imaged 4 orbits in $F475W$ ($g$) and $F625W$ ($r$), 3
in $F775W$ ($i$) and 7 in $F850LP$ ($z$).
We reach $10\sigma$ magnitudes for point sources
(inside a $4\times$FWHM aperture) of $27.5$ in the $g$ band, $27.2$ in
the $r$ and $i$ bands, and $26.7$ in the $z$ band.

\section{Mass modeling}\label{modeling}

The mass model is constructed as follows. We start by assuming an
initial mass model for the cluster, which follows the light of the
cluster-member galaxies. We then identify some sets of multiply-lensed
images, by eye, and force the mass model to reproduce these images, by
changing its parameters. We then use this model to help identify
additional counter images of the multiple images found by eye and also
new sets of multiple images, which are then incorporated into
the model to improve it in an iterative manner.

\subsection{Starting point}

We assign a power law profile to each cluster-member galaxy, assuming
$m \propto L$, making sure that the mass extends way beyond the
boundaries of the ACS image. This is important, since the deflection
field that we wish to construct results from an integration over the
whole mass field. The resulting mass distribution is round on a large
scale, but with small scale structure caused by the galaxies.  We
continue by breaking the mass model into two components. We fit a low
order cubic spline, to get the smooth component, which represents the
dark matter. We than subtract the smooth component from the initial
mass field, and get the galactic residuals, which we refer to as the
lumpy component.  While iterating the model, we allow for the smooth
component to change in shape, using a small number of parameters,
while the lumpy component is allowed to change only in amplitude, 
corresponding approximately to varying the $M/L$ of the galaxy
contribution.

\begin{figure}
 \includegraphics[width=13.5cm]{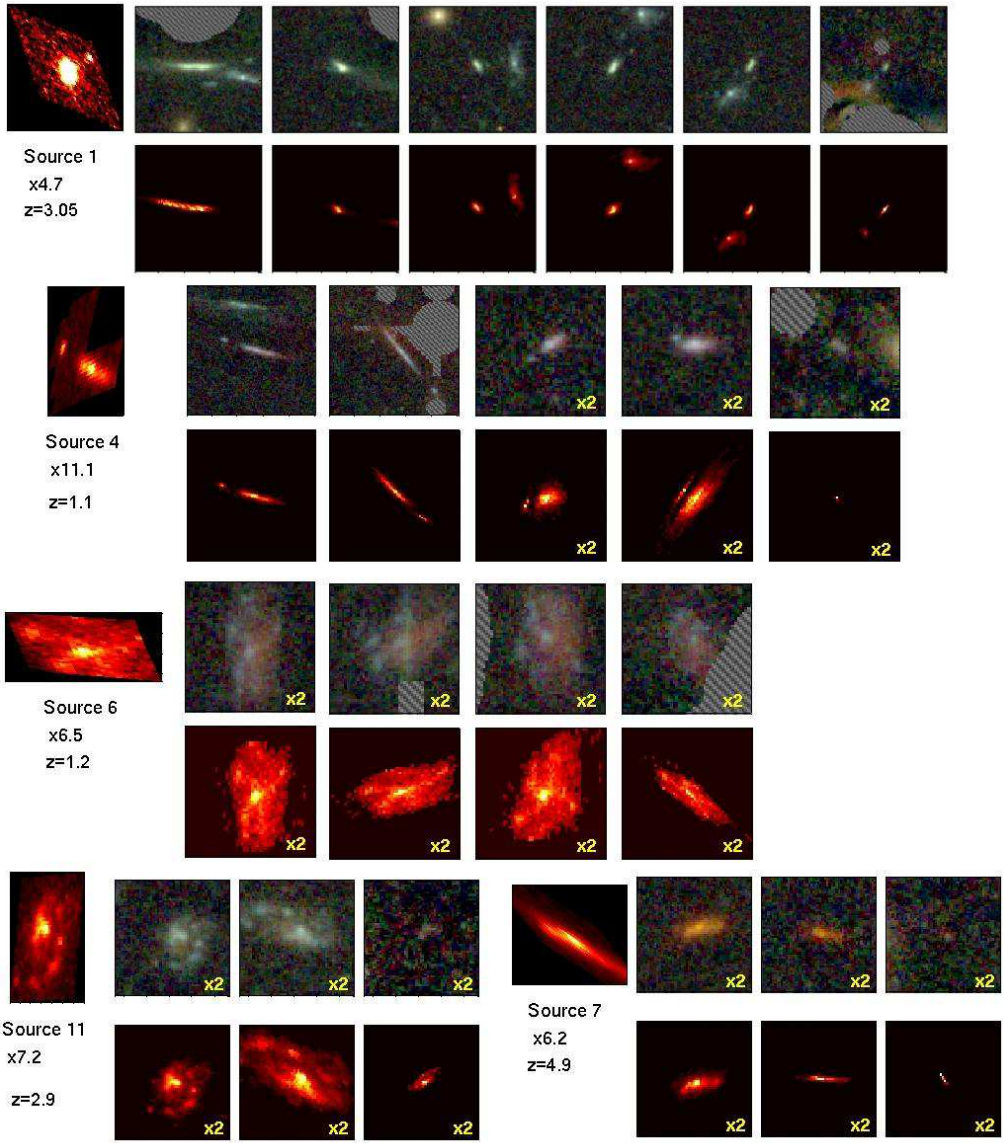}
  \caption{A few examples for multiply lensed galaxies in A1689. For
  each set, we show the observed images in the upper row, while in the
  second row we present the reconstruction of the counter images by
  the best fit model. This comparison is made to demonstrate the
  reliability of the identification of multiple systems.  The
  reconstructed galaxy in the source-plane appears on the left.  Some
  images are magnified by a specified factor, to better demonstrate
  their internal structure.  The redshifts for sources 1 and 7 are
  spectroscopic (Frye, Broadhurst, \& Ben{\'{\i}}tez, 2002), while the
  rest are photometric (Ben{\'i}tez, 2000).}
    \label{fig:stamps}
\end{figure}

\begin{figure}
 \includegraphics[width=13.5cm]{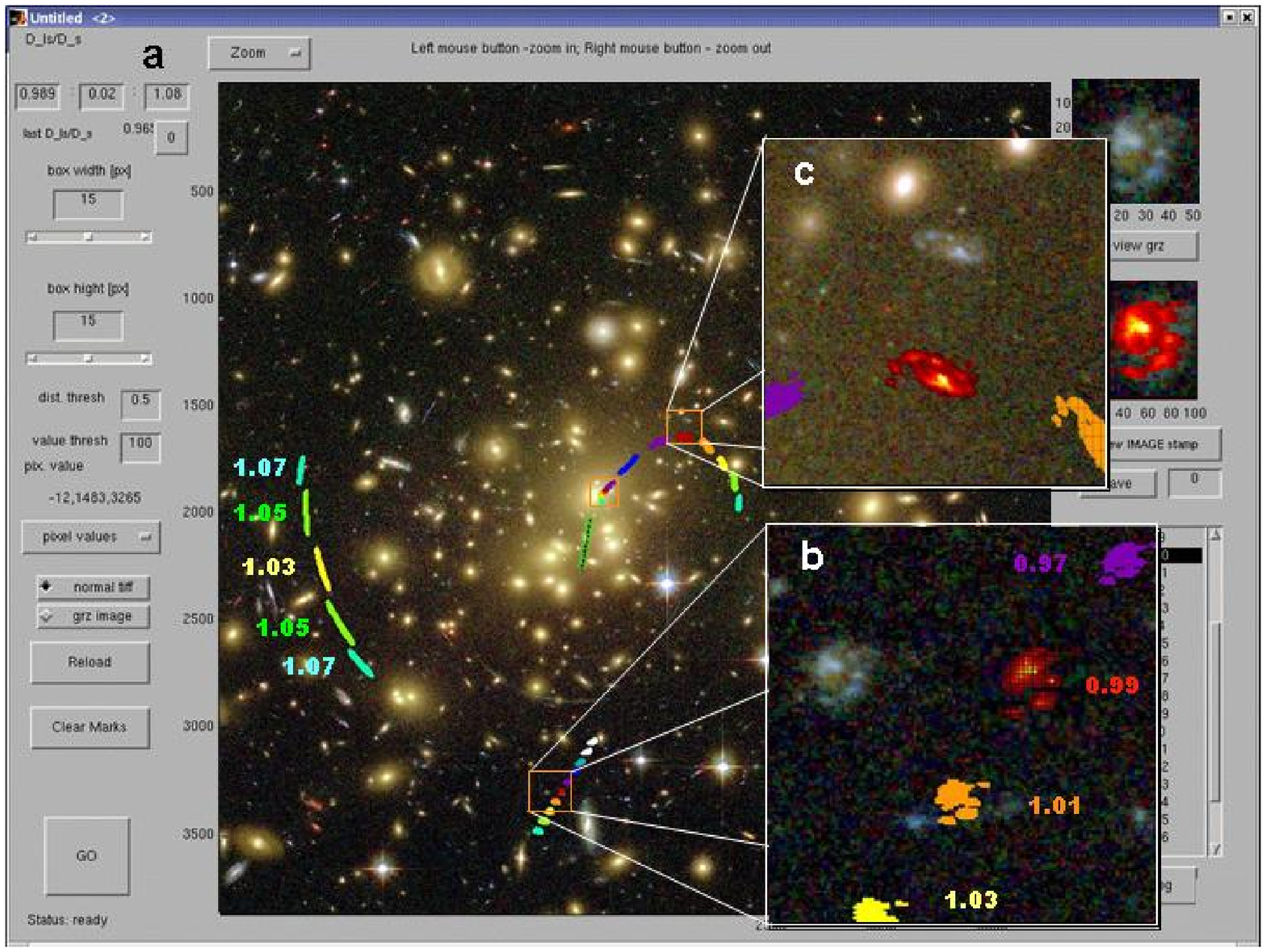}
  \caption{Example of the use of our interactive software, to identify
  multiple image systems.  A range of relative distances for which the
  positions of counter images should be calculates is entered in the left-hand menu ($a$).
  The user clicks on an image of a galaxy, seen here in box $b$, and a
  set of counter-images appear superposed on the cluster image.  The
  colours correspond to different relative distances.  Zooming on area
  $c$, the user can then compare with the data in the vicinity of the
  predicted image.  Note the similarity in shape, structure, parity
  and orientation between the predicted and the observed images.
  Testing for consistency, one can go the other way around and produce
  the counter images of image c, and which appear in box b, clearly
  confirming the identification of this pair of lensed images.  In
  this example, the best fit to the positions corresponds to a
  relative distance scale of 0.993 (where the normalization is is
  fixed to 1.0 at $z=3$).  This galaxy has a photometric redshift of
  $z=2.9$ and may be the highest redshift example of a galaxy showing
  obvious spiral arms.}
    \label{fig:machine}
\end{figure}

\subsection{Multiple image identification}

Some sets of multiply lensed images are detectable by eye.  These are
needed in order to put initial constraints on the model, and improve
its accuracy, before using it to predict additional lensed galaxies.
Some examples of such images are presented in Figure ~\ref{fig:stamps}.
The high spatial resolution and the depth of the ACS data provide
morphological details and internal colour variations which are
repeated in all the counter images of the same source (e.g., sources
4, 6 and 11) so that the identification of counter images is usually
unique. In most of the systems that were discovered by eye, at least
one more additional counter image was predicted by the model and then 
verified visually.

In order to detect less obvious sets of images, we use a new
interactive software, which we wrote in Matlab environment (Fig
\ref{fig:machine}). This software uses the last-iteration deflection
field as an input, and based on it, predicts the locations and
morphology of counter images for a desired range of distance scales ($D_{ls}/D_s$).  The
image identification is done interactively.  Using a graphic user
interface, one can click on a suspected image. The software then
calculates the location and morphology of the galaxy in the source
plane, for a given $D_{ls}/D_s$, using the deflection field and the
values of the pixels which belong to the galaxy. It continues by
scanning the image plane for pixels which are de-lensed to the same
place in the source plane, assigning them with the values of pixels
which are de-lensed to the same location in the source-plane image,
thus conserving surface brightness.

The result of this process, is a set of predicted counter images,
corresponding to the choice of $D_{ls}/D_s$.  The next step, is to
find real images in the vicinity of the predicted images. Once such
image is detected, it can be verified as a counter image, by the
morphology, parity and internal colours.

In order to account for the unknown source distances, one can enter a
desired range of $D_{ls}/D_s$ and the software will produce a sequence of
counter images projected on the observed data frame. Such sequences of
images are seen in Figure~\ref{fig:machine} as extended trails
and are colour-coded to represent different relative distances.

In many cases, the model also predicts very faint counter images,
which lay close to the center of the cluster. These images are harder
to verify, and in fact would be impossible to detect without the help
of the model, as they are out-shined by the light of the cluster
member galaxies and lack internal structure. In such cases, we use an
image in which the bright elliptical galaxies are subtracted. We also
look for sets of nearby galaxies, and make sure that their relative
locations in the innermost parts of the cluster agree with the
predictions of the model.

Whenever several systems of multiply-lensed galaxies are identified,
we fit the model again, 
and continue to look for more systems with the improved model.

\section{Results}
\subsection{Multiple image systems}
\begin{figure}
 \includegraphics[width=13.5cm,height=13.5cm]{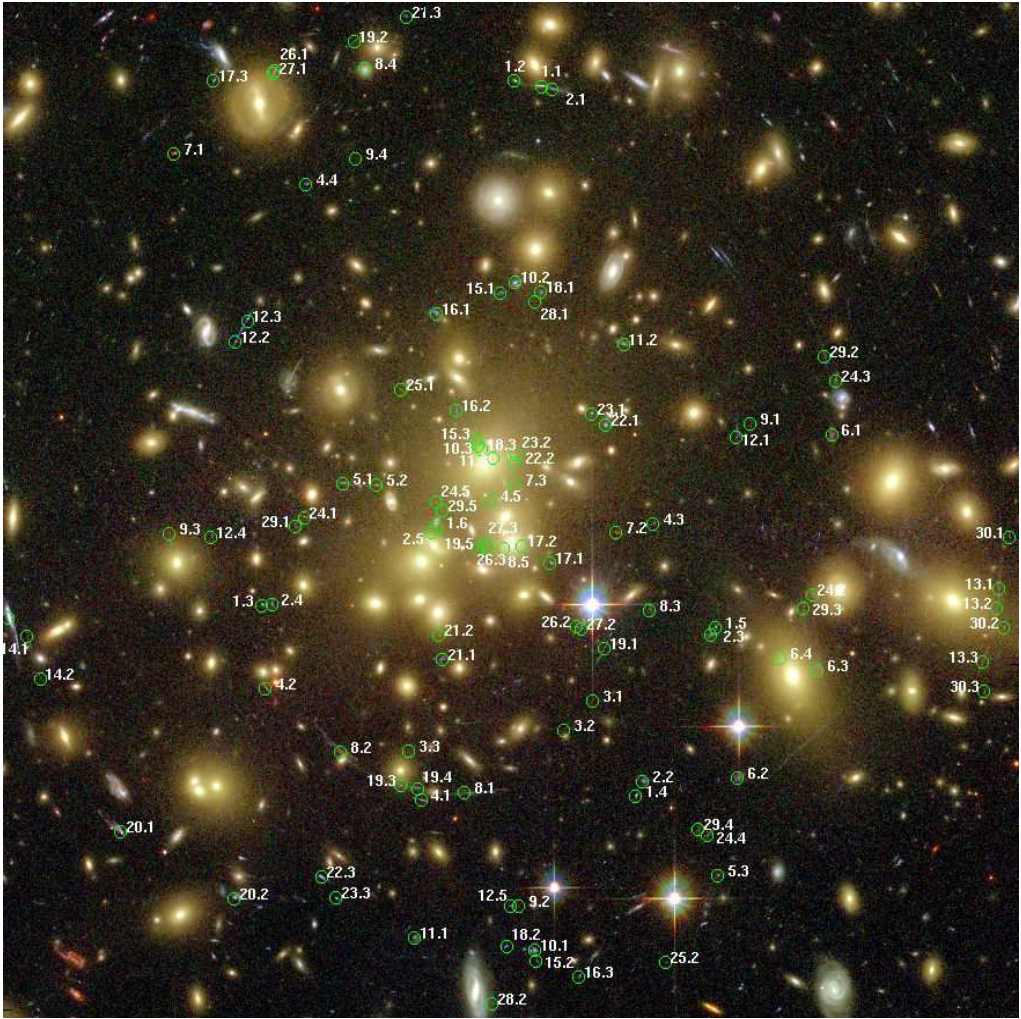}
  \caption{The first 106 images, which were identified with the help of the model
are marked and labeled on the field of the data. The multiple images cover the cluster fairly evenly
including the central region, interior to the radial critical curve.}
    \label{fig:catalog}
\end{figure}

Here we show the first 106 images corresponding to 30 systems of lensed galaxies
(Fig.~\ref{fig:catalog}).  The images are well spread around the
cluster, going all the way into its innermost regions, allowing the
construction of a very detailed mass model. At the center of the
cluster we detect several radial arcs and many sets of images on both
sides of the radial critical curve, tracing its location in great
detail.

Most of those images were detected with the help of the model, in an iterative 
process of improving the model upon the detection of additional images. 
Photometric redshifts are estimated for all images (Ben{\'i}tez, 2000), and usually these
agree well, with the exception of very faint counter-images, or in
cases where there is light contamination from a neighboring cluster
galaxy. The individual multiple-image systems are thoroughly described
in Broadhurst et al. (2004).

\subsection{Conclusions}
We calculate the surface mass distribution in A1689 and deduce the
magnification.  The shape of the mass distribution is approximately
circular in projection and much rounder than the clumpy spatial
distribution of luminous cluster galaxies.  This argues against this
cluster being a lengthly projected filament along the line of sight,
though does not rule out a favorable alignment of a triaxial potential
boosting the observed surface density.

The best fitting mass profile is a good fit to an NFW function, with
relatively high concentration ($C\sim 8$, $r_s\sim 300$kpc/h).  The
magnification derived from this profile reproduces the locations of
the critical curves, and the central magnification is accurately
followed.

We clearly detect the purely geometric increase
of bend-angles with redshift, by comparing the redshift of each source
to the distance scale calculated by the model by minimization in
the image plane.  The dependence on the cosmological parameters is
weak due to the proximity of A1689, $z=0.18$, constraining the locus,
$\Omega_M+\Omega_{\Lambda} \leq 1.2$.  Higher redshift clusters must
be examined in order to constrain the cosmological parameters with
more accuracy.  This consistency with standard cosmology provides
independent support for our model, since the redshift information is
{\it not} used in generating the best fit model.  Similarly, the
relative fluxes of the multiple images are reproduced well by our best
fitting lens model.

The mass to light ratio of A1689 is high,
$M/L(r<250kpc/h)\simeq 400h(M/L_B)_{\odot}$, 30\% larger than any other
well-studied cluster, continuing the general trend of higher M/L
with increasing mass. This large value means that the stellar mass is
a negligible contribution to the overall mass, even in the center
$r<50kpc/h$, where the light is more concentrated than the mass.

\section{Summary} 
We have obtained the highest quality images of a lensing cluster to
date, surpassing previous work both in terms of depth and spatial
resolution and have been rewarded with an order of magnitude increase
in the numbers of multiply lensed galaxies identified around an
individual cluster. This substantial improvement has permitted
detailed modeling of the cluster, which for the first time allows us
to trace a radial critical curve and to measure the mass profile of
the cluster all the way to the center, inside the radial critical
curve, where many small counter-images are found projected on the
center of mass.


\end{document}